# Optical puff mediated laminar-turbulent polarization transition

**Lei Gao[1,*], Tao Zhu[1], Stefan Wabnitz[2,3], Yujia Li[1], Xiao Sheng Tang[1], and Yu Long Cao[1]**

*1 Key Laboratory of Optoelectronic Technology & Systems (Ministry of Education), Chongqing 400044, China*
*2 Dipartimento di Ingegneria dell'Informazione, Università degli Studi di Brescia and INO-CNR, via Branze 38, 25123 Brescia, Italy*
*3 Novosibirsk State University, 1 Pirogova str, Novosibirsk 630090, Russia.*
*\* gaolei@cqu.edu.cn*

**Abstract:** Various physical structures exhibit a fundamentally probabilistic nature over diverse scales in space and time, to the point that the demarcation line between quantum and classic laws gets blurred. Here, we characterize the probability of intermittency in the laminar-turbulence transition of a partially mode-locked fiber laser system, whose degree of coherence is deteriorated by multiple mode mixing. Two competing processes, namely the proliferation and the decay of an optical turbulent puff, determine a critical behavior for the onset of turbulence in such a nonlinear dissipative system. A new kind of optical rogue waves, polarization rogue waves, is introduced at the point of transition to polarization turbulence. The probabilistic description of the puff-mediated laminar-turbulence polarization transition provides an additional degree of freedom for our understanding of the complex physics of lasers.

**References and links**

1. S. Grossmann S, "The onset of shear flow turbulence," Rev. Mod. Phys. 72, 603-618 (2000).
2. D. V. Churkin, S. Sugavanam, N. Tarasov, S. Khorev, S.V. Smirnov, S. M. Kobtsev, and S. K. Turitsyn, "Stochasticity, periodicity and localized light structures in partially mode-locked fiber lasers," Nature Commun. 65, 700 (2015).
3. K. Avila K, D. Moxey, A. de Lozar, M. Avila, D. Barkley, and B. Hof, "The onset of turbulence in pipe flow," Science 333, 192-196 (2011).
4. D. R. Solli, C. Ropers, P. Koonath, B. Jalali, "Optical rogue waves," Nature 450, 1054-1057 (2007).
5. E. Turitsyna, S. V. Smirnov, S. Sugavanam, N. Tarasov, X. Shu, S. A. Babin, E. V. Podivilov, D. V. Churkin, G. Falkovich, and S. K. Turitsyn, "The laminar-turbulent transition in a fiber laser," Nature Photon. 246, 783-786 (2013).
6. J. M. Dudley, F. Dias, M. Erkintalo, G. Genty, "Instabilities, breathers and rogue waves in optics," Nature Photon. 220, 755-764 (2014).
7. D. Nagy, G. Szirmaia, P. Domokos, "Self-organization of a Bose-Einstein condensate in an optical cavity," Eur. Phys. J. D 48, 127-137 (2008).
8. I. Carusotto and C. Ciuti, Quantum fluids of light, Rev. Mod. Phys. 85, 299-366 (2013).
9. M. Sciamanna, K. A. Shore, "Physics and applications of laser diode chaos," Nature Photon. 9, 151-162 (2015).
10. K. Hammani, B. Kibler, C. Finot, A. Picozzi, "Emergence of rogue waves from optical turbulence," Phys. Lett. A 374, 3585-3589 (2010).
11. D. R. Solli, G. Herink, B. Jalali, C. Ropers, "Fluctuations and correlations in modulation instability," Nature Photon. 6, 463-468 (2012).
12. M. Onorato, S. Residori, U. Bortolozzo, A. Montina, F. T. Arecchi, "Rogue waves and their generating mechanisms in different physical contexts," Phys. Rep. 528, 47-89 (2013).
13. P. Walczak, S. Randoux, P. Suret, "Optical rogue waves in integrable turbulence," Phys. Rev. Lett. 114, 143903 (2015).
14. J. P. Eckmann, "Roads to turbulence in dissipative dynamical systems," Rev. Mod. Phys. 53, 643-654 (1981).
15. C. Lecaplain, P. Grelu, J. M. Soto-Crespo, N. Akhmediev, "Dissipative rogue waves generated by chaotic pulse bunching in a mode-locked laser," Phys. Rev. Lett. 108, 233901 (2012).
16. A. Picozzi, J. Garnier, T. Hansson, P. Suret, G. Randoux, G. Millot, and D. N. Christodoulides, "Optical wave turbulence: Towards a unified nonequilibrium thermodynamic formulation of statistical nonlinear optics," Phys. Rep. 542, 1-132 (2014).
17. E. G. Turitsyna, G. Falkovich, V. K. Mezentsev, S. K. Turitsyn, "Optical turbulence and spectral condensate in long-fiber lasers," Phys. Rev. A 80, 031804R (2009).

18. L. Gao, T. Zhu, S. Wabnitz, M. Liu, W. Huang, "Coherence loss of partially mode-locked fiber laser," Sci. Rep. 6, 24995 (2016).
19. M. Haeltermann, S. Trillo, S. Wabnitz, "Polarization multistability and instability in a nonlinear dispersive ring cavity," J. Opt. Soc. A. B 11, 446-456 (1994).
20. R. Christian, K. Michael, C. Lucia, W. Benjamin, R. Piotr, M. Clerici, Y. Jestin, M. Ferrera, M. Peccianti, A. Pasquazi, B. E. Little, S. T. Chu, D. J. Moss, and R. Morandotti, "Cross-polarized photon pair generation and bi-chromatically pumped optical parametric oscillation on a chip," Nature Commun. 6, 8236 (2015).
21. V. Kalashnikov, S. V. Sergeyev, G. Jacobsen, S. Popov, S. K. Turitsyn, "Multi-scale polarisation phenomena," Light Science & Applications 5, e16011 (2016).
22. S. V. Sergeyev, C. Mou, E. G. Turitsyna, A. Rozhin, S. K. Turitsyn, "Spiral attractor created by vector solitons," Light Science & Applications 3, e131 (2014).
23. R. Nissim, A. Pejkic, E. Myslivets, B. P. Kuo, N. Alic, and S. Radic, "Ultrafast optical control by few photons in engineered fiber," Science 345, 417-419 (2014).
24. K. Goda, and B. Jalali, "Dispersive Fourier transformation for fast continuous single-shot measurements," Nature Photonics 7, 102–112 (2013).
25. M. Sano, K. Tamai, "A universal transition to turbulence in channel flow," Nature Phys. 12, 249-254 (2016).
26. G. Herink, B. Jalali, C. Ropers, D. R. Solli, "Resolving the build-up of femtosecond mode-locking with single-shot spectroscopy at 90 MHz frame rate," Nature Photon. 10, 321-326 (2016).
27. Z. Liu, S. Zhang, F. W. Wise, "Rogue waves in a normal-dispersion fiber laser," Opt. Lett. 40, 1366-1369 (2015).

## 1. Introduction

The intermittent transition from laminar to turbulent wave propagation has been observed in many physical systems, ranging from galactic down to microscopic scales such as spiral galaxies, fluid motions in atmosphere and oceans, blood vessels, and even financial markets [1-3]. Non-trivial statistics are promoted irreversibly by the combined action of advection and mixing processes, leading to a departure from the random probability density functions (PDF) of Gaussian behavior, or power-law distributions [4-6]. The study of light propagating in nonlinear optical fibers provides an excellent testbed for advancing our understanding of the laminar-turbulent transition, based on a probabilistic description. Both highly coherent states, such as Bose-Einstein condensate-like [7] or nonlinear photon superfluids [8], and low coherent states such as deterministic chaos have been investigated in theory and identified in lightwave experiments [9]. The degree of coherence of a laser system is controlled by the total energy of the light field, and it has been shown to spontaneously evolve from a fully deterministic state into an incoherent turbulent state [5]. During the transition to turbulence, short-lived giant optical waves with high-amplitudes appear and disappear erratically without leaving a trace. These studies open up new avenues for advancing our understanding of the still hotly debated mechanisms for the generation of optical rogue waves associated with the elevated tails of corresponding L-shaped PDFs [4,6,10,11].

The universality of the turbulence scenario is persistent both in conservative and in dissipative dynamics. Yet, a general description of the process leading to turbulence is still missing. So far, wave turbulence theory has essentially been developed to describe weakly interacting turbulence, while the strong turbulence regime is associated with near-integrable systems, which in the integrable limit exhibit analytical solutions such as solitons and breathers [12,13]. In dissipative laser systems, complex wave structures are formed, which cannot be deterministically predicted from given initial conditions [14]. Moreover, the increase of the number of degree of freedom, such as the mutual interacting longitudinal modes with periodic boundary conditions, leads to complex and chaotic phenomena in laser cavities, for example, strange attractors, period doubling, and positive Lyapunov exponents [15-17].

To fully describe the transition to turbulence, all dimensions of the optical field, namely, amplitude, phase, and polarization need a proper consideration. Similar to pipe flow, it has been proved that with an increase of pump power or cavity length, turbulence arises due to the spatial breakdown of coherence [5]. This light-fluid analogy shares the same probabilistic nature. Yet, more work is required to characterize the turbulence spread or decay, and to reveal the universal mechanisms of turbulence development in laser systems. In particular, methods leading to the identification, prediction, and control of optical rogue waves both in their spatiotemporal regime and even in their polarization evolution regime need to be further investigated.

In this article, we experimentally study the probabilistic nature of intermittency in the laminar-turbulence transition of a partially mode-locked fiber laser (PML), where longitudinal modes interact via multiple wave mixing. We discover the occurrence of two competing processes: in analogy with spatial puffs in pipe flow [3], we unveil the proliferation and the decay of an optical turbulent puff, which determines the critical point leading to the onset of polarization turbulence in a fiber laser system. Moreover, we experimentally characterize a new kind of vector rogue waves, characterized in terms of their state of polarization (SOP). We show that, whenever the coherence of the system decreases, the SOP bifurcates on the Poincaré sphere. The polarization pattern formation process is based on multiple wave mixing occurring among the parametric instability (PI) sidebands [18-23]. A laser system operating with pure PI is laminar: its behavior in the temporal, spectral, and the polarization regimes remains highly coherent. However, when the pump power is further increased, cascaded four-wave mixing processes lead chaotic fluctuations of the laser amplitude, phase, and SOP. As a result, the coherent laser system behavior is broken, until it evolves into fully developed turbulence.

## 2. EXPERIMENTAL SETUP

The fiber cavity shown in Fig. 1 contains 1 m erbium-doped fiber (EDF, Liekki ER 80-8/125) pumped by two 976 nm continuous wave (CW) lasers through two wavelength division multiplexers (WDMs), 14.5 m of single mode fiber (SMF), 19.5 m of dispersion compensating fiber (DCF), a polarization independent optical isolator (ISO), a polarization controller (PC), and an optical coupler (OC), where 10% is utilized for the output. The dispersions of the EDF, DCF, and SMF are 15.7, -38, and 18 ps/(nm*km), respectively, resulting into a net normal dispersion of 0.737 $ps^2$. We insert a saturable absorber (SA) fabricated by filling reduced graphene oxide (RGO) flakes into cladding holes of a photonic crystal fiber (PCF), to provide enough nonlinear saturable absorption for mode-locking. The RGO was prepared by reducing chemical oxidation of graphite, which has been intensively investigated for laser mode-locking, due to the Pauli blocking. We produce a transparent RGO solution through centrifuging an ultrasound-treated RGO-N,N-dimethylformamide solution, and fill the cladding holes of the PCF with the RGO solution based on Siphon Effect[2]. After drying in a vacuum chamber, a 2 cm PCF is spliced between two sections of SMFs with a total insertion loss of 5.8 dB, which involves a splice loss of 3 dB and an absorption loss of 2.8 dB. For the Grapefruit-shaped PCF used here, the light energy flow through the RGO is only $1/10^7$ of the light passing through the inner core [17]. Therefore, the thermal damage threshold of the SA can be substantially increased. The modulation depth of the fabricated SA is about 24%. Due to two-dimensional structure of RGO, the polarization-dependent loss of the SA is ~3 dB.

The laser output is detected by two kinds of photodetectors (PD1, 350 MHz; PD2, 50 GHz). The bandwidth of the oscilloscope connecting PD1 is 1 GHz: the recorded pulse amplitude is proportional to the pulse energy, since the rise time of PD1 is much larger than the pulse duration. The single-shot spectra are detected by dispersive Fourier transformation (DFT) [11,24]: periodic signals are stretched by means of a normal dispersive fiber with a dispersion of 300 $ps^2$ for frequency-to-time transformation, and subsequently fed to a 50 GHz PD2 connected to a real-time oscilloscope with bandwidth of 20 GHz. The resolution of the

DFT is less than 0.2 nm. An optical spectrum analyzer (OSA) with a resolution of 0.02 nm is utilized for single-shot spectra with spectra averaged over multiple passes in the cavity. Temporal features are detected by an autocorrelator with a delay resolution of 6 fs. The state of polarization (SOP) of the partially mode-locked laser (PML) is measured by means of a high-speed polarization state analyzer (PSA). Due to larger insertion loss of the tunable filter (TF), an erbium-doped fiber amplifier (EDFA) is utilized for signal amplification. The bandwidth of the TF is 1 nm.

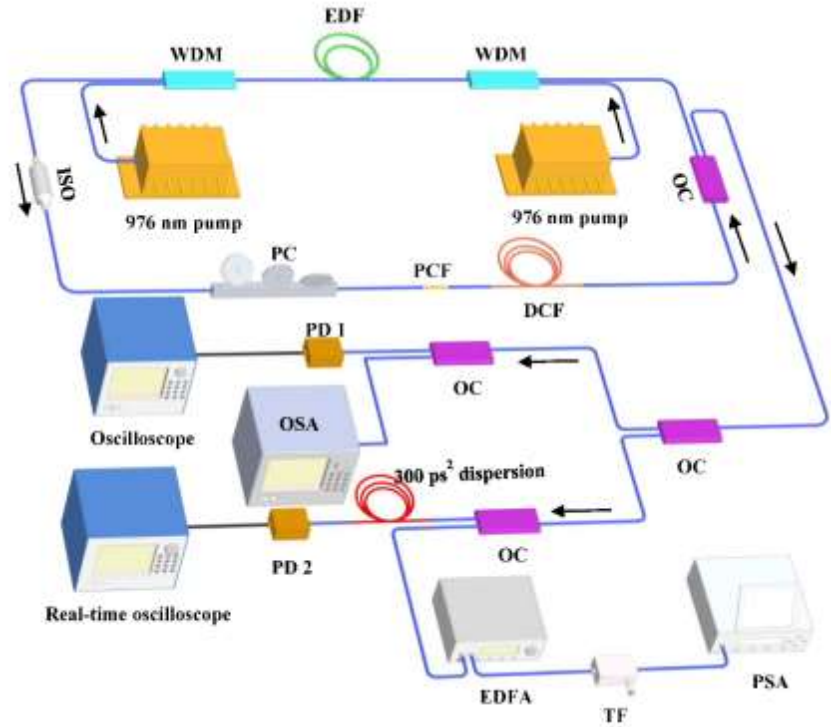

Fig. 1. Schematic of our fiber laser cavity and measurement methods. (a) EDF, erbium-doped fiber (LiekkiER 80-8/125); WDM, wavelength division multiplexer; ISO, polarization independent optical isolator; PC, polarization controller; DCF, dispersion compensation fiber; OC, optical coupler; OSA, optical spectrum analyzer; PD, photo-detector; EDFA, erbium-doped fiber amplifier; TF, tunable filter; PSA, polarization state analyzer.

## 3. Results and discussion

PML was achieved with proper PC detuning when the intra-cavity EDF was pumped by two counter-propagating CW lasers with power of 400 mW at 976 nm [17], corresponding to a total pump power of 800 mW. Figure 2 depicts the typical output when the laser system is operating in a stable PML mode. Owing to the low bandwidth (350 MHz) of PD1, its response time is much longer than the typical pulse duration of PML, therefore the corresponding oscilloscope trace in Fig. 2(a) does not reveal significant intra-pulse intensity fluctuations. The autocorrelation trace in Fig. 2(b) exhibits a pedestal with a full width at half maximum (FWHM) of 30.5 ps, plus a coherent peak with a duration of about 416 fs. This kind of structure is a result of the autocorrelation of a series of temporal pulses exhibiting strongly fluctuating structures. The magnitude of fluctuations within each pulse can be easily identified from their corresponding single-shot spectra as detected by DFT in Fig.2(c), which are invisible in the averaged optical spectrum. The FWHM of the optical spectrum as detected by a conventional optical spectrum analyzer is 16 nm, which is consistent with the corresponding single-shot spectra detected by DFT.

Based on the analogies between the dynamics of fluids and lasers, as they share similar nonlinear mode coupling behaviors, in the presence of stochastic boundary conditions, we introduce the concept of optical turbulence puff, and anticipate a critical behavior in the formation of optical turbulence [3,25]. Puffs exhibit turbulent localized patches either in the temporal (in optics) or the spatial (in fluids) domains, are embedded in laminar backgrounds. Our optical puffs are transient waves whenever the pump power is relatively low (corresponding to a Reynolds number below 2000 in fluids). They proliferate and decay at different speeds in a memoryless manner, according to the power in the cavity. Fig. 3 depicts the formation of optical puffs in the temporal domain at various pump power levels, where the average power and optical spectra are measured with the help of the DFT as a function of the discrete time variable, $t=N*t_{rt}$ ($t_{rt}$ is the round-trip time of the fiber ring cavity).

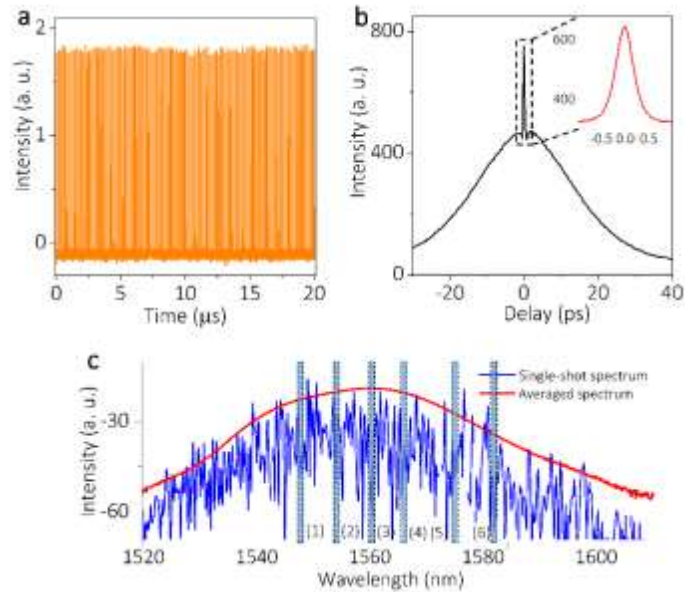

Fig. 2. Typical outputs for PML. (a) Oscilloscope trace. (b) Autocorrelation trace (inset is the coherent peak in a larger scale). (c) Example of single-shot spectrum together with the averaged spectrum.

As shown by Fig.3(a,b), our experiments show that, when the pump power $P$ is below 750 mW, optical puffs decay with a speed that is faster than the speed associated with their proliferation. Above this power value, Fig.3(c,d) shows that a turbulent spectrum starts to be self-sustained. As can be seen in Fig.3(a), in the front-leading part of a puff appearing in the proliferating regime, the laser system comprises a gradually decreasing CW emission, accompanied by sub-picosecond-scale temporal fluctuations within each pulse (as revealed by the narrow autocorreleation peak in fig.3(b)), whose amplitude is exponentially growing with the number of round-trips. Cavity gain dynamics as determined by the superposition of CWs and pulses leads to an oscillatory intracavity power behavior vs. the number of round trips (see Fig.3(a)).

The whole process is reminiscent of the dynamics observed in titanium-doped sapphire mode-locked lasers [26], but it occurs over much reduced time scales (the build-up time of femtosecond lasers requires $10^6$ round trips, while it takes several hundred round trips for building PMLs). The optical spectra in this region are narrow and smooth. The oscillatory behavior of intracavity power in Fig.3(a) is attributed to dynamic gain competition. Due to the nonlinear saturable absorption of the SA and the periodically alternating signs of fiber dispersion in the cavity, the laser spectra swiftly broaden in the turbulent regime through the interplay of DFWM, FWM, and stimulated Raman scattering [17].

To ascertain the stochastic nature of puff-mediated transition to optical turbulence, we studied its statistics, in analogy with the case of hydrodynamics [3]. Let $P_{proliferating}(P, t)$ be the probability that an optical puff starts to disappear before time $t$ at pump power $P$,

$$\text{PR}_{proliferating}(P,t) = \exp((-t-t_0)/\tau(P)) \qquad (1)$$

where $t_0$ is the puff formation time (see panel b of Fig.3, for P=650 mW), and $\tau(P)$ is a pump-dependent characteristic lifetime of a specific fiber ring cavity. The dependence of probability P on the number of round-trips for different pump power levels is presented in Fig.4(a). Each point here is obtained from 5000 measurements.

To eliminate any possible influence of trigger from the oscilloscope, we ignore the data in the first puff in any long temporal segment. The experimental probability distributions in Fig. 4(a) confirm well the hypothesis of an exponential distribution of optical puff decays as described by Eq.(1), thus indicating that the formation of turbulence is a memoryless stochastic process with characteristic time $\tau(P)$.

The corresponding characteristic time $\tau(P)$ was calculated by fitting our experimental data with the theoretical formula (1). Figure 4(b) shows that, in the build-up of an optical turbulent puff, the characteristic time $\tau(P)$ decreases exponentially with $P$. This is reasonable, since the fiber laser system requires a smaller number of round-trips to accumulate a given nonlinear phase for larger $P$ values. This non-diverging behavior is universal in shear flow or Couette flow turbulence, indicating that the turbulent optical puff will decay and the whole system will re-laminarize whenever the intracavity power remains relatively low. The probability of forming an optical puff is thus only determined by pump power $P$.

The critical behavior of optical puff generation in the transition to laser turbulence can thus be identified by determining the characteristic times of puff proliferation and decay. This criticality is more apparent in the single-shot spectra in Fig. 3(e), where proliferation time and decay time are almost the same. The critical point defines the boundary for the outweight of the two competing processes (namely, pump proliferation and decay), below which the system is unable to sustain turbulence and is relaminarized, and vice versa. When the pump power is large enough, more optical puffs are generated, and subsequently they merge and couple with each other. As a result, the laser system evolves into full turbulence (see Fig.3(c,d), and Fig.3(f), for P=750 mW).

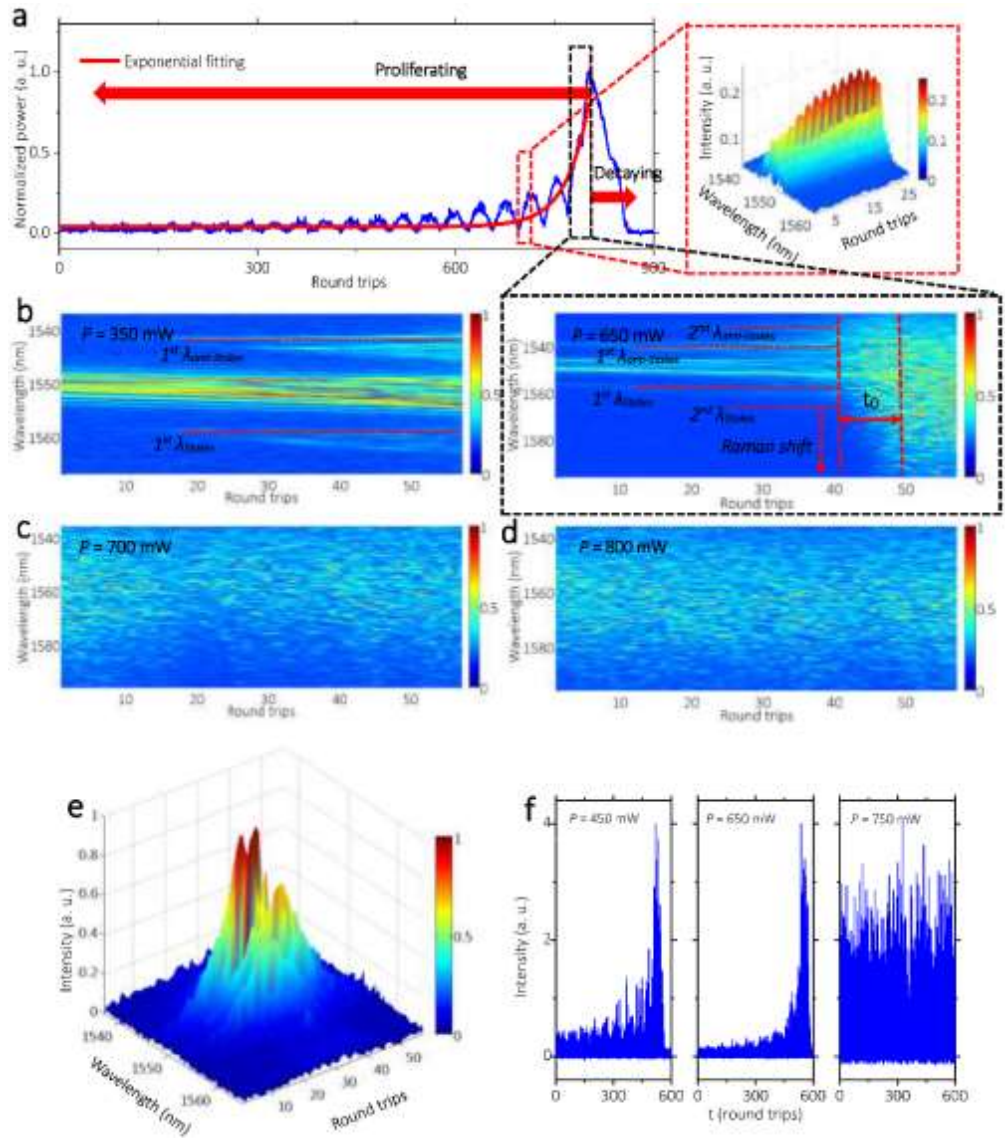

Fig. 3. The build-up of PML as the formation of optical turbulent puff. (a) Normalized intracavity power in forming an optical puff, which demarcates two regimes, namely the proliferation phase (as fitted exponentially by means of a solid red line) and the decay phase. About 850 round trips are required for the optical puff to proliferate from pure noise to a maximum intensity value. Next, the puff decays within a time frame, which is much smaller than the proliferation time. The time $t_0$ for the turbulence formation as shown by single-shot spectra in panel (e) is about 10 $t_{rt}$. Whenever the total energy is insufficient to sustain mode-locking, pulse relaminarizes. (b)-(d) single-shot spectra for various pump powers. The corresponding single-shot spectrum appears as stochastically varying at each round trip. (e) Single-shot spectra with almost identical puff proliferation and decay times, corresponding to the critical point for transition to turbulence, for $P$=650 mW. (f) temporal outputs for various pump powers.

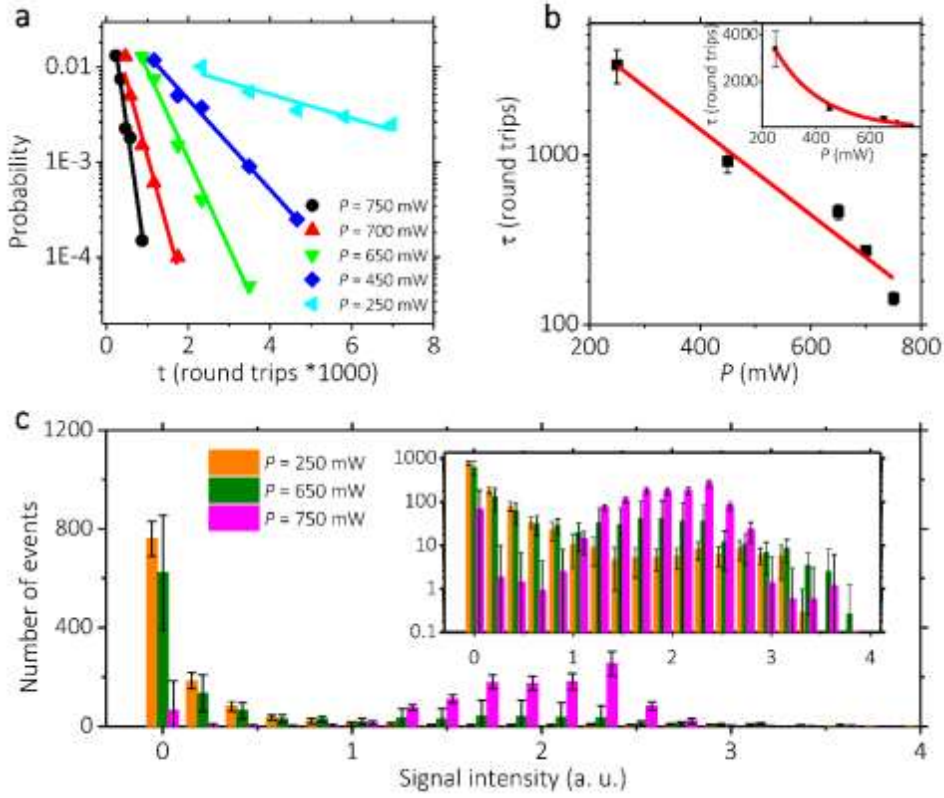

Fig. 4. Statistical properties of temporal laser emission. (a) Measured probability of optical puffs starting to decay up to time $t$. All data are fitted with the expression $\exp[-(t - t_0)/\tau(P)]$, as illustrated by the straight lines. The formation time of an optical puff, $t_0$, is about 10trt as deduced from experimental data. (b) Variation of characteristic time $\tau$ with pump power $P$ (inset in linear scale). This was extracted by exponentially fitting data in (a). Error bars represent absolute errors based on experimental uncertainties. The error is relatively high for $P$ larger than 600 mW, since the corresponding $\tau$ is smaller. This error may have a contribution from the oscilloscope trigger error, and the difficulty to distinguish two adjacent optical puffs when $P$ is high. (c) Histograms of the temporal trace (inset in log scale), showing that the signal intensity distribution exhibits a transition from laminar to turbulent state. Optical rogue waves can be well distinguished from stochastic events.

It is during this transition from laminar to turbulent states, that optical rogue waves are identified in the temporal domain, and described by their own probability distribution. As shown in Fig. 4(c), the histogram of the temporal trace for P at 250 mW already exhibits a non-Gaussian, L-shaped distribution. As the pump power grows larger, the histogram expands for higher signal intensities, corresponding to the generation of optical puffs. When the pump power is raised up to about 750 mW, the histogram reveals an isolated large amplitude peak, corresponding to the formation of rogue waves.

In the analysis of large skewed probability distributions, we should keep in mind that the temporal pulse trace is a combined superposition of all lasing frequencies. The photodetector with a bandwidth of 350 MHz is unable to identify the fine structure of each pulse. Therefore, we filtered the output spectra as shown in Fig. 3 by means of a wavelength tunable filter with a bandwidth of 1 nm: the resulting scattering plots for six wavelengths are shown in Fig. 5, showing the fluctuations of the laser output energy between two successive round-trips. It is clear that, far away from the center laser line (1558.43 nm), the scattering of the temporal traces becomes more pronounced. Scattering in red-shifted regions is exacerbated by the presence of stimulated Raman scattering. A similar phenomenon was reported when studying rogue wave generation in a normal dispersion fiber laser based on the DFT [27]. The experimental results of Fig.4(c) indicate that optical mediated transition to turbulence in our laser is associated with the generation of optical rogue waves.

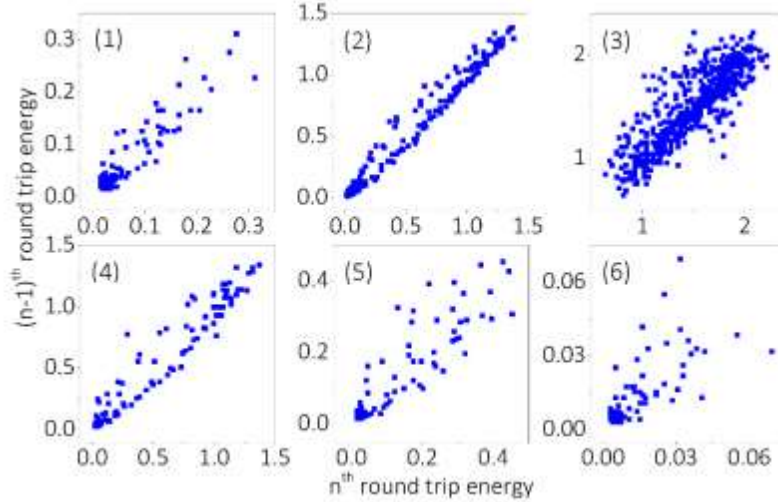

Fig. 5. Scattering plots for filtered PML at various wavelengths. The wavelength are 1547.6 nm, 1556.1 nm, 1561.2 nm, 1565 nm, 1574.1 nm, 1582.7 nm, respectively, which are indicated by (1), (2), (3), (4), (5), (6) in Fig. 3(c).

Another inherently fundamental parameter of fiber laser emission, its SOP, has received relatively less attention in the literature. In particular, the generation of turbulence in the laser SOP has not yet been fully demonstrated. In order to observe the laminar-turbulent transition of the SOP in PML, we measured the SOP of each filtered wavelength at different pump powers. Figures 6(a)-(c) show that when optical puffs are formed, the corresponding SOPs for wavelengths far away from the laser center line bifurcate into a cross-like shape on the Poincaré sphere. This is in marked contrast with the single fixed point that is observed for pump powers $P$ smaller than 80 mW, in which case no puff can be detected. As earlier discussed in Ref.[18], multiple vector wave mixing processes generate new frequencies with separately evolving output SOP azimuth and ellipticity angles, resulting in perpendicular lines aligning with either a meridian or a parallel curve on the Poincaré sphere, respectively. The two SOP curves along the parallel of the Poincaré sphere bifurcate, similar to what is observed in the formation of chaos [9]. Here, the operating state of the laser can be regarded as laminar, since the laser system still exhibits a high degree of coherence.

The onset of SOP turbulence is accompanied by the loss of temporal coherence. As shown by Fig.6(a-c), obtained by filtering the wavelength $\lambda=1556.1$ nm, which is close to the center of the laser line, irregular polarization states located outside of the main polarization directions emerge when the pump power $P$ exceeds 250 mW. This scattering of SOPs gets more pronounced when the intracavity power is further increased up to P=650 mW. This result seems natural when we consider the usual route for chaos, resulting from a cascade of successive period-doubling bifurcations. FWM cascades with nonlinear chaotic phase-matching conditions originating from nonlinear net phase shifts in the resonator. Whenever the pump power is larger than 600 mW, cascaded FWM leads to a fully developed turbulent evolution, and the SOPs of filtered wavelengths appears as totally random.

The irregular polarization state of laser emission is associated with the emergence of a new kind of optical rogue waves in PML, namely, optical polarization rogue waves. At variance with temporal or spatial rogue waves, polarization rogue waves have a vector nature. We may introduce the following method for their characterization. The SOP on the Poincaré sphere is expressed in terms of the Stokes vector $\hat{S}=(s_1, s_2, s_3)$. The relative distance between the SOPs of N different polarized waves is defined by

$$r = \sum_{\substack{m,n=1 \\ m \neq n}}^{N} \left| \hat{S}_m - \hat{S}_n \right| = \sum_{\substack{m,n=1 \\ m \neq n}}^{N} \sqrt{(s_{m1} - s_{n1})^2 + (s_{m2} - s_{n2})^2 + (s_{m3} - s_{n3})^2} \qquad (2)$$

The PDFs of the distance $r$ between the various SOPs for different pump powers and wavelengths are depicted in Figs. 6(d,e,f). A mere cross-like bifurcation of the SOP (like that in Fig.6(a), for a pump power P=250 mW) results in exponential distributions. On the other hand, Figs.6(d,e,f) show that the fully developed turbulent regime for the SOP is accompanied by PDFs for the distance $r$ which resembles a lognormal distribution. Moreover, the width of the turbulent PDFs is strongly wavelength dependent, being minimal close to the center of the laser line. ($\lambda$=1558,3 nm).

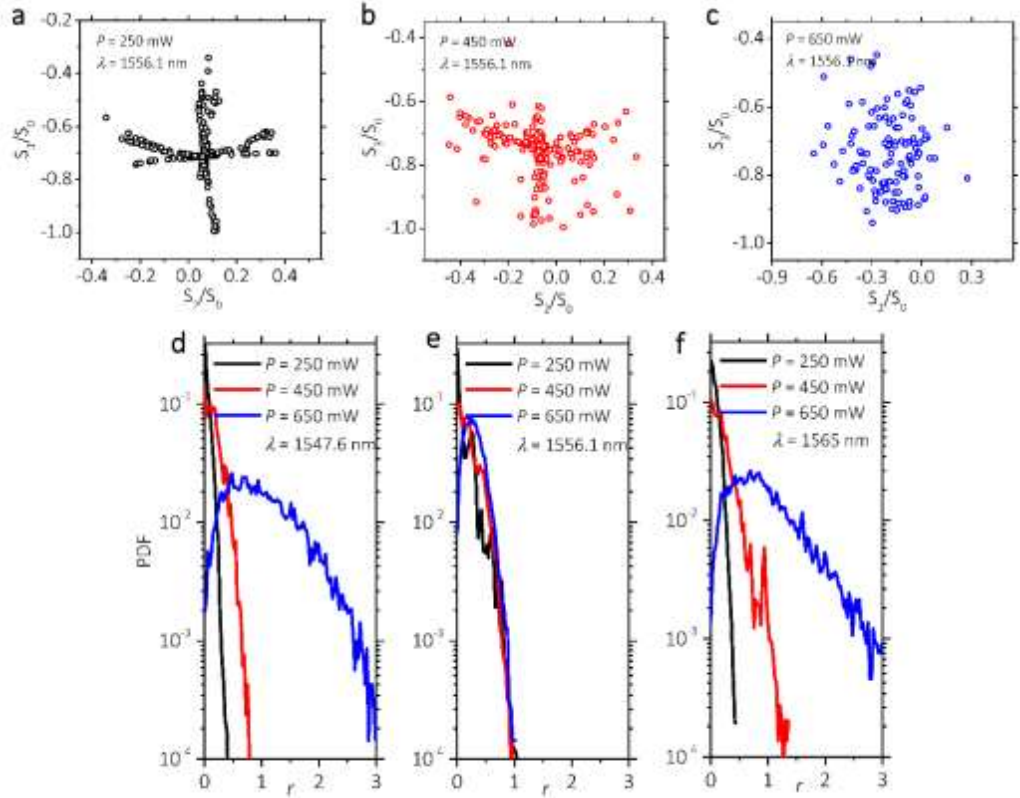

Fig. 6. Optical polarization rogue waves in the polarization laminar-turbulent transition. (a)-(c), Normalized Stokes parameters $s_2$, $s_3$ for three different filtered wavelength at different pump powers. Irregular points with distance far away from others are observed along with unpredictable probabilities and positions, indicating the emergence optical polarization rogue waves. (d)-(f), Logarithm of the PDF of the distance between each polarization point

corresponding to cases in (a)-(c). The PDF evolves into a nearly lognormal distribution when more optical turbulent puffs are formed, where the SOP is fully randomized.

## 4. Conclusions

We experimentally investigated the probabilistic nature of intermittency in the laminar-turbulence polarization transition in a partially mode-locked fiber laser. We introduced the concept of optical turbulent puff, and identified a corresponding new critical behavior for the onset of turbulence, which is associated with the loss of temporal coherence. A similar universal turbulence scenario was also found for the corresponding SOPs evolution, where the presence of polarization rogue waves was identified. Those results reveal that the coherence loss of a laser system may be attributed to the stochastic mixing of its vector modes, which form complex turbulent structures. Our results provide a new insight into the still debated topic of the mechanisms for rogue wave generation. This route for the transition between laminar and turbulence dynamics unveils a new method for the control and exploitation of non-equilibrium optical systems.

## Funding

This work was supported by the Key Research and Development Project of Ministry of Science and Technology (2016YFC0801200), the Natural Science Foundation of China (61705023, 61635004, 61405020, 61520106012), the National Postdoctoral Program for Innovative Talents (BX201600200), the General Financial Grant from the China Postdoctoral Science Foundation (2017M610589), the Postdoctoral Science Foundation of Chongqing (Xm2017047), the Science Foundation of Chongqing (CSTC2017JCYJA0651), the Science Fund for Distinguished Young Scholars of Chongqing (CSTC2014JCYJJQ40002) and the Fundamental Research Funds for the Central Universities (106112017CDJXY120004). Stefan Wabnitz acknowledges support by the Russian Ministry of Science and Education (14. Y26.31.0017), the European Union's Horizon 2020 research, and innovation programme under the Marie Skłodowska-Curie grant agreement No 691051

## Acknowledgement

Beneficial discussions with Sergey Turitsyn are gratefully acknowledged.